\documentstyle[12pt]{article}

\newcommand{\bc}{\begin{center}}
\newcommand{\ec}{\end{center}}

\begin{document}

\title{On the relation of Matrix theory and Maldacena conjecture}
\author{Pedro J. Silva\thanks{email:pedro.silva@ncl.ac.uk  psilva@delta.ft.uam.es}\\ Physics department, University of Necastle, UK NE1 7RU\\
Depto de F\'{\i}sica Te\'orica, Universidad Autonoma de Madrid,\\Cantoblanco 28049, Madrid, Spain}
\maketitle

\begin{abstract}
We report a sign that M(atrix) theory conjecture and the Maldacena
conjecture for the case of D0-branes are compatible. Furthermore Maldacena
point of view implies a restriction of range of validity in the DLCQ version
of M(atrix) theory. The analysis is based on the uplift of type IIA
supersymetric solution in the Maldacena approach to eleven dimensions, using
a boost as a main tool. The relation is explored on both, IMF and DLCF
versions of M(atrix) theory.\newline
\end{abstract}

\input epsf \vspace{24pt}

\pagebreak \newpage

\section{Introduction}

In the last three years, a multitude of dramatic developments in string
theory have appeared as a direct consequence of the so called string
dualities. Between this conjectures a mayor step forward was taken by the
M(atrix) theory proposal of Banks, Fischler, Shenker and Suskind (BFSS) \cite
{bfss}. This conjecture states that M-theory in the infinite momentum frame
(IMF) is equivalent to a $U(N)$ matrix quantum mechanics, in the limit $N
\rightarrow \infty$. Also there is another version due to Susskind \cite
{sus1}, that identified the discrete lightcone frame (DLCF) of M-theory with
$U(N)$ matrix quantum mechanics. This later theory is understood as the low
energy limit of the worldvolume theory that describes the dynamics of $N$
D0-branes (The proposal that related super Yang Mills theory in 10
dimensions reduce to (p+1) dimensions, with the low energy world volume
theory of Dp-branes was done by Witten \cite{wit1}). Precisely the D0-branes
are the key objects for the above conjecture once is noted that they
represent the only possible candidates with momentum on the compactified
eleven dimension, where we are thinking on type IIA string theory as the
result of a compactification on the eleventh dimension of M-theory.

A few mouths ago Maldacena formulated a new conjecture \cite{mal1}, relating
the theory of the world volume of $N$ D3-branes for big $N$, with type IIB
supergravity on the product of five dimensional Anti-desiter spacetime and
compact manifold, that in the maximal symmetric case is a five sphere. A
presice recipe for this duality is formulated on \cite{wit2}. Later on this
conjecture was extend to other values of $p\;\; (0,1,2,4,5,6)$ \cite{mal2}.
Referent to the $p=0$ case, a studie on the conformal structure not
explicity showed on this model, appere recentely, renforcing the conjecture
\cite{jev1}.

Although the nature and interpretation of this to conjectures (BFSS and
Maldacena) seems to be in principle different, it is clear that both ideas
have an amazing resamblens. Note that the worldvolume theory on the low
energy regime of Dp-branes are invoque, the large limit for N is crucial for
both conjectures, both theories give and M-theory result, and depend heavily
on the BPS character of the Dp-branes.

On this letter we report an interpretation for the relation between both
approach. In doing so, the Maldacena conjecture is set for ($p=0$), then the
specific character of the D0-branes as Kaluza-Klein modes is crucial. Also a
verification for the validity of the relation is given by showing that
Maldacena analysis support the exact limit of IMF of M(atrix) Theory. The
relation with DLCF is more precise and we get a better idea of why these
similarities, as well as a one to one correspondence for any $N$, between
both approaches.

\section{Maldacena conjecture and $Ads_2 \otimes S_8$}

Following Maldacena \cite{mal2}, we write the exact form of the metric
describing D0-branes on type IIA supergravity in the limit of low energy.
\begin{eqnarray}
\alpha ^{\prime}\rightarrow 0\;\;,\;\;g_s \rightarrow {\alpha ^{\prime}}%
^{3/2}\;\;\,\;\;r^s \rightarrow \alpha ^{\prime}
\end{eqnarray}
such that the following definitions remind constant,
\[
g_{YM}=(2\pi)^{-1}g_s^{1/2}\alpha^{\prime-3/4}\;\;\;U=\alpha^{\prime}/r_s
\]
where $\alpha^{\prime}$ is the square of the plank length, $g_s$ is the
string coupling constant, $r_s$ represents spacelike directions
perpendicular to the brane, and $g_{YM}$ is the Yang Mills coupling constant
on the effective worldvolume theory of the brane. The subindex $s$ refers to
the ten dimensional nature of these definitions in the so called string
metric, that is where the Ramon-Ramon field is not couple to the dilaton.
The explicit form of the metric is,
\begin{eqnarray}
ds^{2}=\alpha^{\prime}\left[-\frac{U^{7/2}}{d_o^{1/2}\sqrt{N}g_{YM}}dt^2+%
\frac{d_o^{1/2}\sqrt{N}g_{YM}}{U^{7/2}} dU^2+\frac{d_o^{1/2}\sqrt{N}g_{YM}}{%
U^{3/2}}d\Omega_{8}\right] \\
e^{\phi}=(g_{YM}2\pi)^2\left(\frac{d_o g_{YM}^2 N}{U^7}\right)^{3/4} \\
A_0=\frac{U^7}{d_o g_{YM}^2N(g_{YM}2\pi)^{8/3}}
\end{eqnarray}

From the point of view of Maldacena conjecture, the theory on the world
volume on the Do-brane is dual to supergravity on the above solution, when
the energy scale ``$U$'' satisfies the following inequallyties,
\begin{equation}
N^{1/7}<<\frac{U}{g_{YM}^{2/3}}<<N^{1/3}  \label{eq:b1}
\end{equation}
where the upper bound correspond to the small curvature constraint and the
lower bound to a small effective string coupling.

To have a connection to M-theory theory, we should recover the full eleven
dimension, and somehow represent strong coupling regime.

Let's now uplift this solution of type IIA supergravity to one of eleven
dimension supergravity. This is easy, as by construction Type IIA is the
dimensional reduction of eleven dimensional supergravity (for example see
\cite{ste1}). Note that for the case of D0-branes we only need to consider
the Einstein-Hilbert term of eleven dimensional supergravity. We use the
following standard notation, where we have ignored the constant in front of
the action.
\begin{equation}
S^{11}=\int {dx^{11}\sqrt{G}R(G)}
\end{equation}
where the metric is of the form,
\begin{eqnarray}
ds^{2} &=&e^{\frac{-2}{3}\phi }ds_{10}^{2}+e^{\frac{4}{3}\phi }(dz-A_{\mu
}dx^{\mu })^{2} \\
e^{\phi } &=&g_{s}f^{3/4}\;\;,\;\;A_{0}=\frac{-1}{f}\;\;,\;\;z\in (0,1)
\nonumber
\end{eqnarray}
Note that the 2-form obtained after this reduction is indeed the source of
RR field of the D0-branes. To be able of taking the low energy limit to rich
the Maldacena framework we consider the following change of variables on the
eleventh coordinate sector,
\begin{equation}
dx^{11}=g_{s}^{-1/3}(dz+\frac{1}{f})
\end{equation}
therefore after taking the low energy limit ($\alpha ^{\prime }\rightarrow
\infty $), we obtain the eleven dimensional metric,
\begin{eqnarray}
ds^{2} &=&\alpha ^{\prime }\left[ -\frac{U^{7}}{(d_{o}g_{YM}^{2})(g_{YM}2\pi
)^{4/3}N}dt^{2}+\frac{(d_{o}g_{YM}^{2})(g_{YM}2\pi )^{4}N}{U^{7}}%
dx_{11}^{2}+\right.   \nonumber \\
&&\left. +\frac{1}{(g_{YM}2\pi )^{2}}(du^{2}+U^{2}d\Omega _{8})\right]
\nonumber \\
e^{\phi } &=&(g_{YM}2\pi )^{2}\left( \frac{d_{o}g_{YM}^{2}}{U^{7}}\right)
^{3/4}  \label{eq:m1}
\end{eqnarray}
note that the definition of $dx^{11}$ is well behave on the low energy limit
taken. Up to this moment we have only reconstruct the eleven dimensional
point of view of the type IIA supergravity solution describing D0-branes.
The eleventh dimension is currently compactified, and the radio of curvature
is very small, as we really are in a ten dimensional ``observable
supergravity''.

On the M(atrix) side, we count with two different versions for the
conjecture, one related to IMF and the other to DLQF. As these two versions
are not exactly the same, we most consider different approach to find a
connection with the Maldacena conjecture.

The key idea on the forthcoming analysis comes from the simple observation
that the Maldacena conjecture defines a relation between the theory on the
worldvolumme of $N$ D0-branes and M/superstring theory on a given
background, while M(atrix) theory defines another relation on the same
worldvolume theory to M-theory on a specific frame, either IMF or DLCF.
Therefore by assuming both conjectures correct, we should get a relation
between M-theory on IMF/DLCF and the uplift metric showed above.

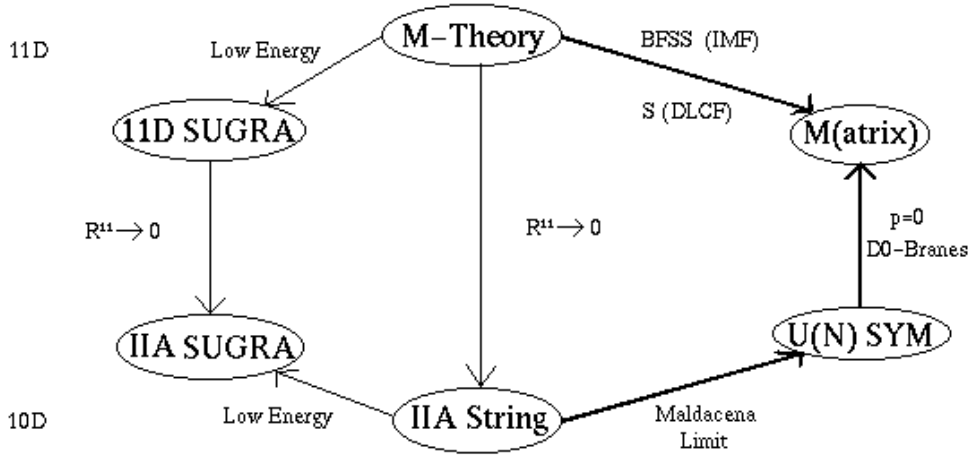
\begin{center}
\vskip .5truecm \hskip 0truecm
\vbox{
        \epsfysize=6truecm
        \epsffile{mfig.epsi}
}
\end{center}

\section{DLCF and Type IIA D0-branes Solutions}

To make contact with DLQF, we need to related a observer in a frame with a
spacelike compact dimension with another observer in a frame this time with
a lightlike compact direction. Therefore boosting to this new frame should
be analyzed in detail.

First consider the to rewrite the metric of equation (\ref{eq:m1}) on $%
\alpha ^{\prime }$ units as,
\begin{equation}
ds_{11}^{2}=\left[ -\frac{U^{7}}{AN}dt^{2}+\frac{BN}{U^{7}}dx_{11}^{2}+\frac{%
1}{C}dU^{2}+\frac{U^{2}}{C}d\Omega _{8}\right]
\end{equation}
where $x^{11}\in (0,(g_{YM}2\pi )^{-2/3})$, $A=(d_{o}g_{YM}^{2})(g_{YM}2\pi
)^{4/3}$, $B=(d_{o}g_{YM}^{2})(g_{YM}2\pi )^{4}$ and $C=(g_{YM}2\pi )^{2}$.
Therefore the Radio of the eleventh compactified dimension is obtain as $%
R_{0}=(g{YM}2\pi )^{4/3}\sqrt{\frac{Nd_{o}g{YM}^{2}}{U^{7}}}$. This metric
stands as the tensor product of two spaces, a sphere of radius $U$, and at
each point of a given value of $U$ a Lorentzian cylinder. We will consider
the above metric on trajectories of fixed value for $U$, therefore, at a
given energy scale for the worldvolume theory of the D0-branes. The same
kind of analysis is considered when investigate properties of a given
metric. Next we consider a boost along the compactified eleventh dimension,
following Seiberg approach \cite{sei1} to DLCQ. The rapidity is choosing to
be,
\begin{equation}
\beta =\frac{R}{\sqrt{R^{2}+2R_{0}^{2}}}
\end{equation}
Initially we have the following identification,
\begin{equation}
\left(
\begin{array}{c}
t \\
x^{11}
\end{array}
\right) \sim \left(
\begin{array}{c}
t \\
x^{11}
\end{array}
\right) +\left(
\begin{array}{c}
0 \\
R_{0}
\end{array}
\right)
\end{equation}
in the new frame we have the identification,
\begin{equation}
\left(
\begin{array}{c}
x^{+} \\
x^{-}
\end{array}
\right) \sim \left(
\begin{array}{c}
x^{+} \\
x^{-}
\end{array}
\right) +\left(
\begin{array}{c}
\frac{R_{0}}{2R} \\
-R+\frac{R_{0}}{2R}
\end{array}
\right)
\end{equation}
where $x^{+}$,$x^{-}$ are the usual lightcone coordinates in the boosted
frame. As usual we get the DLCF in the limit of $\beta \rightarrow 1$.

As a next step, we identify the above M-theory with a new M-theory, on the
DLCF with momentum $\bar{P}_{-}$ and plank length $\bar{\alpha}^{\prime }$.
The relation on the momentum sector is given by,
\begin{equation}
\overline{P}_{-}\overline{\alpha }^{^{\prime }}=P_{11}
\end{equation}
Using the fact that $x^{-}$ is compact, we have a precise form for the
momentum on the $x^{-}$ component,
\begin{equation}
\bar{P}_{-}=\frac{\bar{N}}{R}
\end{equation}
but as $\bar{N}$ counts the numbers of D0-branes, $N=\bar{N}$, therefore we
found that the new M-Theory is not anymore at low energies, because on units
of $\alpha ^{\prime }$, in the new M-theory we have,
\begin{equation}
\bar{\alpha}^{\prime }=\frac{R}{R_{0}}\rightarrow \infty
\end{equation}
As always, The lightcone frame is obtained by letting $(N,R)\rightarrow
\infty $ while keeping $P_{-}$ constant.

Note that in the above relation we never were forced to trespass the
constraint on the energy scale $U$, only in the process of
decompactification, such a violation would occur. Therefore we are ready to
say, as a conclusion of the above analysis that `` the DLQF of M(atrix)
theory reproduce only the M-theory behavior at a given range of energy scale
on the worldvolume theory of $N$ D0-branes, and that the presice scale of
energy is deffined by the the inequality of equation \ref{eq:b1}

Basically we have identified the M(atrix) DLCF version of M-theory with an
uplifted 11D supergravity solution on the Maldacena conjecture, such that
for fixed $N$ we have a constraints on the range of values of $U$.

\section{IMF and Type IIA D0-branes Solutions}

To make contact with the IMF version we should consider boosting the
solution, in the eleventh direction, in the limit of ($N,R^s_{11}$)$%
\rightarrow\infty$ such that $P_{11}$ goes to infinity. As in our metric the
radius is a function of ($N,U$) there is no guaranty that the above limit is
include as a possibility. Let us consider explicitly this limit,
\begin{equation}
P^s_{11}=\frac{N}{R^s_{11}}=\frac{N^{1/4}U^{21/4}}{(2\pi)^2 g_{YM}^{7/2}}
\label{eq:p11}
\end{equation}
For $U$ within the Maldacena limits, we obtain the followings bounds for $%
P^s_{11}$,
\begin{equation}
N<<P^s_{11}<<N^2
\end{equation}
This doesn't take us to the IMF prescribe in M(atrix) theory as $R^S_{11}$
remains small when $N\rightarrow\infty$. To obtain the desirable IMF
prescription, we are oblige to go beyond the Maldacena limits. We found
that, in order to get, infinite momentum and infinite radio simultaneously,
the variable $U$ should satisfy the inequalities,
\begin{equation}
\frac{1}{N^{1/21}}<<\frac{U}{g_{YM}^{2/3}}<<N^{1/7}
\end{equation}

To obtain the IMF we are force to decompactify the eleventh dimension, this
is translate in a new range of validity for the values of $U$, so that in
the limit
\[
N\rightarrow \infty \;\;\hbox{where}\;\;(P_{11}^{s},R_{11}^{s})\rightarrow
\infty
\]
we recover the full axis of positive values of $U$ i.e. $(0,\infty )$. Note
that we go onto a non-perturbative hight energy regime, M-theory.

As a summary we have,

\begin{itemize}
\item  $N$ (in the Matrix theory) measures the momenta in units of $%
1/R_{11}^{s}$

\item  $N$ is proportional to the radius of AdS and also to the String
coupling constant.
\end{itemize}

The relation then goes at follows: The eleventh dimensional momentum $P_{11}$
is constraint by the geometry, the IMF limit would correspond to a
decompactification of the 11 dimension, therefore from the supergravity
point of view, you are going to a strong coupling, but you can do it in such
a way that curvature is still small, and as a bonus the whole range of
possible energies on the super Yang Mills theory is regain.

\section{Digression}

It would be very interesting to consider similar analysis for the other
values of $p$ by the use of  T-duality on the D-branes solutionos. In
particular the case $p=4$ would be of specific importance. The relation
between the generalized superconformal algebra found on D0-branes \cite{jev1}
and the $SO(2,4)$ algebra of the $AdS_{5}$ should be investigated. This will
allow to study duality relations between correlators by an approach a la
Witten.

\vspace{12pt} {\bf Acknowledgements}

We thanks the theoretical Physics group of IFT of UAM for interesting
conversations on the subject. This work was supported by the Venezuelan
Government.

\bibliographystyle{unsrt}
\bibliography{mm}

\end{document}